\documentclass[a4paper]{jpconf} 
\usepackage{graphicx} 
\usepackage{amssymb,amsmath,cite} 
\usepackage{slashed} 
\usepackage[dvipsnames]{xcolor} 
\def\lamb#1#2{$^{#1}_{\Lambda}${#2}}

\def\lamblamb#1#2{$^{~~#1}_{\Lambda\Lambda}${#2}} 
\def\lamlam#1#2{$^{~#1}_{\Lambda\Lambda}${#2}}

\newcommand{\nopieft}{\mbox{$\slashed{\pi}$EFT}}

\newcommand{\be}{\begin{equation}} 
\newcommand{\ee}{\end{equation}}

\newcommand{\rvec}{{\vec{r}}}

\begin{document} 

\title{Recent progress on hypernuclei} 
\author{Avraham Gal} 
\address{Racah Institute of Physics, The Hebrew University, 
91904 Jerusalem, Israel} 
\ead{avragal@savion.huji.ac.il} 

\begin{abstract} 
Some of last year's progress made in hypernuclear physics is reviewed 
as follows: (i) resolving the \lamb{5}{He} overbinding problem in 
single-$\Lambda$ hypernuclei~\cite{cbg18}; (ii) arguing that the onset 
of binding double-$\Lambda$ hypernuclei is most likely at $A$=5, with the 
neutral systems \lamblamb{3}{n} and \lamblamb{4}{n} unbound by a large 
margin~\cite{csbgm19}; and (iii) revising the calculated value of the loosely 
bound \lamb{3}{H} lifetime to a level of $\sim$20\% shorter than the free 
$\Lambda$ lifetime~\cite{GG18}, given recent claims from relativistic heavy 
ion experiments that $\tau$(\lamb{3}{H}) is shorter than $\tau_{\Lambda}$ by 
as much as $\approx$(30$\pm$8)\%. Also discussed briefly in this context is 
the lifetime expected for the questionable \lamb{3}{n} hypernucleus. 
\end{abstract}

\section{Introduction} 
\label{sec:intro} 

Single- and double-$\Lambda$ hypernuclei provide a unique extension of 
nuclear physics into strange hadronic matter~\cite{ghm16}. Experimental 
data on $\Lambda$ and $\Lambda\Lambda$ hypernuclei are unfortunately poorer 
both in quantity and quality than the data available on normal nuclei. 
Nevertheless, the few dozen $\Lambda$ separation energies $B_\Lambda$ of 
single-$\Lambda$ hypernuclei ($_{\Lambda}^{\rm A}$Z) determined across the 
periodic table from $A$=3 to 208, and the three $\Lambda\Lambda$ hypernuclei 
($^{~\rm A}_{\Lambda\Lambda}$Z) firmly established so far, provide a useful 
testground for the role of strangeness in dense hadronic matter, say in 
neutron star matter. Particularly meaningful tests of hyperon-nucleon and 
hyperon-hyperon strong-interaction models are possible in light $\Lambda$ 
and $\Lambda\Lambda$ hypernuclei, $A\leq 6$, which the three topics reviewed 
below are concerned with. Before focusing on each topic separately, I discuss 
briefly in the next section the pionless EFT ({\nopieft}) framework which is 
used in two of these topics.

\section{{\nopieft} methodology}
\label{sec:EFT} 

\begin{figure}[htb] 
\begin{center} 
\includegraphics[width=0.62\textwidth]{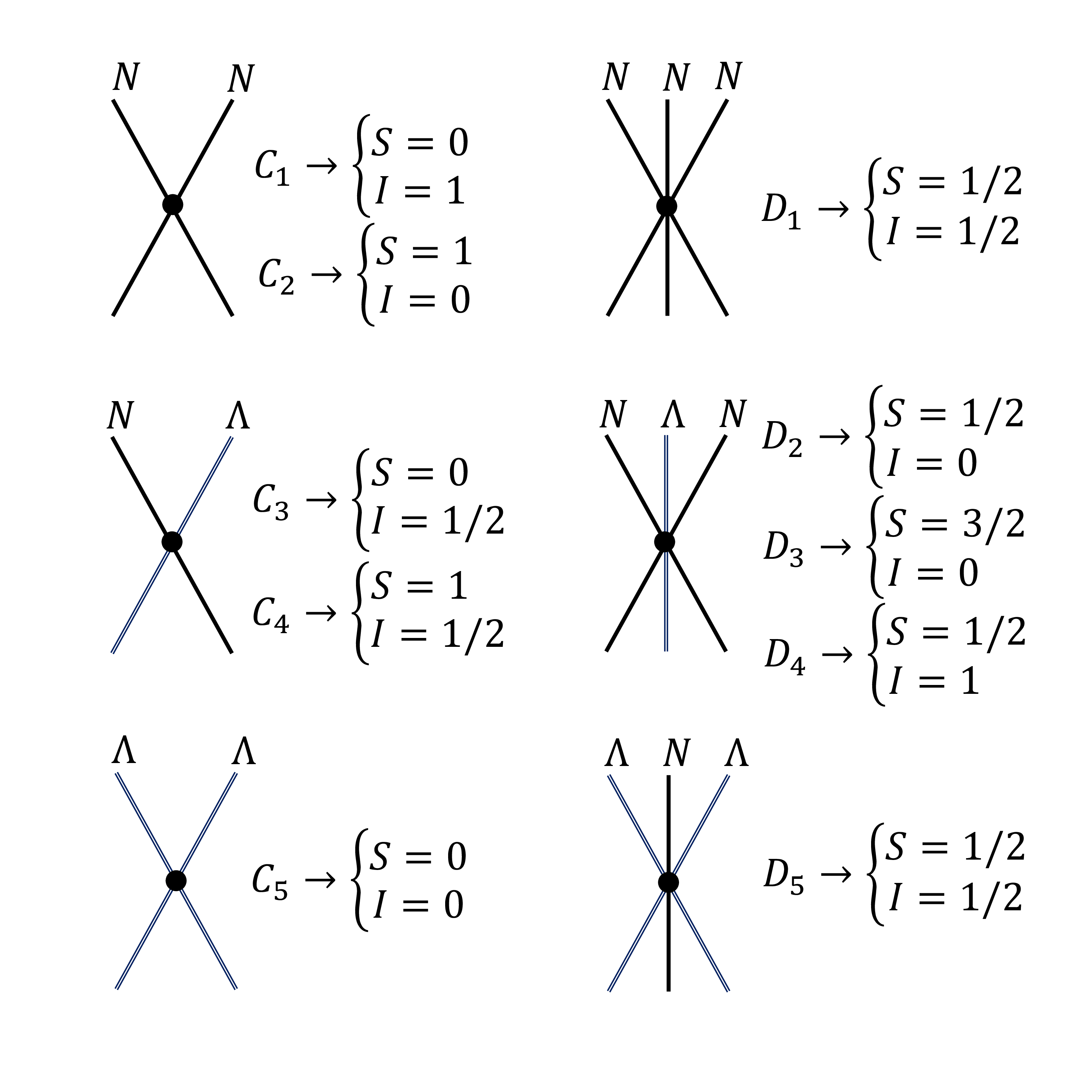} 
\caption{Diagrammatic presentation of two-body (left) and three-body
(right) contact terms, and their associated LEC input ($C_1,\ldots,C_5$)
\& ($D_1,\ldots,D_5$) to a LO {\nopieft} calculation of light nuclei
(upper), $\Lambda$ hypernuclei (middle) and $\Lambda\Lambda$ hypernuclei
(lower), with values of spin $S$ and isospin $I$ corresponding to $s$-wave
configurations. Figure adapted from Ref.~\cite{csbgm19}.} 
\label{fig:diag} 
\end{center} 
\end{figure} 

The leading-order (LO) {\nopieft} interaction for aggregates of nucleons and 
$\Lambda$ hyperons consists of two-body and three-body $s$-wave contact 
interaction terms shown diagrammatically, together with the corresponding 
low-energy constants (LECs) listed alongside, in Fig.~\ref{fig:diag}. Each LEC 
is labelled by the total Pauli-spin $S$ and isospin $I$ involved. Further 
contact terms, such as a three-body $\Lambda\Lambda\Lambda$ term, appear 
only at subleading orders. These two-body and three-body contact interaction 
terms give rise to two-body and three-body potentials
\begin{equation} 
V_2 = \sum_{IS}\,C_{\lambda}^{IS} \sum_{i<j} {\cal P}_{IS}(ij)
          \delta_\lambda(\rvec_{ij}),\,\,\,\,\,  
V_3 = \sum_{\alpha IS}D_{\alpha\lambda}^{IS} \sum_{i<j<k}
{\cal Q}_{IS}(ijk)\left(\sum_{\rm cyc}\delta_\lambda(\rvec_{ij})
\delta_\lambda(\rvec_{jk})\right), 
\label{eq:V} 
\end{equation}
where ${\cal P}_{IS}$ project on $s$-wave $NN,\Lambda N,\Lambda\Lambda$ 
pairs with isospin $I$ and spin $S$ values associated with two-body LECs 
in Fig.~\ref{fig:diag}. These LECs are fitted to low-energy two-body 
observables, e.g., to the corresponding $NN,\Lambda N,\Lambda\Lambda$ 
scattering lengths. Similarly, ${\cal Q}_{IS}$ project on $NNN$, $NN\Lambda$ 
and $\Lambda\Lambda N$ $s$-wave triplets with isospin $I$ and spin $S$ values 
associated in Fig.~\ref{fig:diag} with three-body LECs which are fitted 
to given binding energies. The subscript $\alpha$ distinguishes between the 
two $IS=\frac{1}{2}\frac{1}{2}$ $NNN$ and $\Lambda\Lambda N$ triplets marked 
in the figure. The subscript $\lambda$ attached to $C^{IS}$ and $D^{IS}$ in 
Eq.~(\ref{eq:V}) stands for a momentum cutoff introduced in a Gaussian form 
to regularize the zero-range contact terms:
\begin{equation}
\delta_\lambda(\rvec)=\left(\frac{\lambda}{2\sqrt{\pi}}\right)^3\,
\exp \left(-{\frac{\lambda^2}{4}}\rvec^{\,2}\right),   
\label{eq:gaussian} 
\end{equation}
thereby smearing a zero-range (in the limit $\lambda\to\infty$) Dirac
$\delta^{(3)}(\rvec)$ contact term over distances~$\sim\lambda^{-1}$. The
cutoff parameter $\lambda$ may be viewed as a scale parameter with respect 
to typical values of momenta $Q$. To make observables cutoff independent, 
the LECs must be properly renormalized. Truncating {\nopieft} at LO and 
using values of $\lambda$ higher than the breakup scale of the theory 
which is of order 2$m_{\pi}$ for the isoscalar $\Lambda$ hyperon, observables 
acquire a residual dependence $O(Q/\lambda)$ which diminishes with increasing 
$\lambda$. Using such two-body $V_2$ and three-body $V_3$ regularized 
contact interaction terms, the $A$-body Schr\"{o}dinger equation was 
solved by expanding the wave function $\Psi$ in a correlated Gaussian 
basis, using the stochastic variational method (SVM). 

Few-body {\nopieft} calculations were first reported for nucleons 
in Refs.~\cite{kol99,bhk00} and recently extended to lattice 
nuclei~\cite{bcg15,kbg15,clp17,kpd17}. Past hypernuclear applications 
are limited to \lamb{3}{Z}~\cite{hammer02,ando15,hammer19} and to $A$=4,6 
$\Lambda$-$\Lambda$-core three-body calculations~\cite{ando14a,ando14b}. 
The calculations reviewed below are the first systematic single- and 
double-$\Lambda$ hypernuclear studies covering the full nuclear $s$ shell.

\section{Overbinding of \lamb{5}{He}}
\label{sec:L5He} 

The overbinding of \lamb{5}{He} upon using fitted two-body $\Lambda N$ 
interactions, even when adding $\Lambda NN$ terms owing to $\Sigma$ hyperon 
excitation, was first recognized and stated clearly in a 1972 landmark paper 
by Dalitz {\it et al.}~\cite{dht72}. There, as well as in recent LO chiral 
effective field theory ($\chi$EFT) calculations~\cite{wr18}, the $\Lambda$ 
separation energy $B_{\Lambda}$(\lamb{5}{He}) comes out as large as 6~MeV, 
well above the value $B_{\Lambda}^{\rm exp}$(\lamb{5}{He})=3.12$\pm$0.02~MeV, 
as demonstrated in the first two main rows of Table~\ref{tab:over}. 
No truly ab-initio calculations of $B_{\Lambda}$(\lamb{5}{He}) using 
next-to-leading-order (NLO) $\chi$EFT interactions have ever been reported. 
Comprehensive NLO calculations for $A$=3,4 hypernuclei have recently been 
published~\cite{hmn19}, with results listed in the last two rows of the table. 
The cutoffs 500, 650~MeV chosen for the 2013 and 2019 versions, respectively, 
are motivated by looking for those cutoff values that correspond to $\Lambda$ 
well-depth values in the range $D_{\Lambda}=28-30$~MeV. Given that such NLO 
versions fit the low-energy $\Lambda$p cross sections~\cite{Alex68} better 
than the LO model~\cite{phm06} does, it is puzzling why the latest 2019 
NLO version does so poorly for the $A$=4 hypernuclei, definitely worse than 
the LO calculation~\cite{gg16} does. 

\begin{table}[!h] 
\caption{Ground-state $\Lambda$ separation energies $B_{\Lambda}$ and 
excitation energies $E_x$ (in MeV) from several few-body calculations 
of $s$-shell $\Lambda$ hypernuclei, see text. Charge symmetry breaking 
is included in the $_\Lambda^4$He results from Ref.~\cite{gg16}.} 
\begin{center} 
\begin{tabular}{ccccc} 
\br 
& $B_{\Lambda}(_\Lambda^3$H) & $B_{\Lambda}(_\Lambda^4$He$_{\rm g.s.}$) 
& $E_x(_\Lambda^4$He$_{\rm exc.}$) & $B_{\Lambda}(_\Lambda^5$He)  \\ 
Exp. & 0.13(5)~\cite{davis05} & 2.39(3)~\cite{davis05} & 
1.406(3)~\cite{E13} & 3.12(2)~\cite{davis05}  \\ 
\mr 
$\chi$EFT(LO$_{600}$) & 0.11(1)~\cite{wirth14} & 2.444~\cite{gg16} & 
1.278~\cite{gg16} & 5.82(2)~\cite{wr18}  \\
$\chi$EFT(LO$_{700}$) & -- & 2.423~\cite{gg16} & 1.941~\cite{gg16} & 
4.43(2)~\cite{wr18}  \\
$\chi$EFT(NLO13$_{500}$) & 0.135~\cite{hmn19} & 1.705~\cite{hmn19} & 
0.915~\cite{hmn19} & -- \\ 
$\chi$EFT(NLO19$_{650}$) & 0.095~\cite{hmn19} & 1.530~\cite{hmn19} & 
0.614~\cite{hmn19} & -- \\ 
\br 
\end{tabular} 
\label{tab:over} 
\end{center} 
\end{table} 

Here I review a rather successful attempt to resolve the \lamb{5}{He} 
overbinding problem within the simpler EFT approach of pionless EFT 
(\nopieft), limited at LO to nucleons and $\Lambda$-hyperons degrees 
of freedom, by means of precise SVM calculations of $s$-shell 
hypernuclei~\cite{cbg18}. Note that the long-range $\Lambda N\to\Sigma N$ 
one-pion exchange (OPE) transition followed by an equally long-range 
$\Sigma N\to\Lambda N$ OPE transition is dominated by its central 
$S\to D\to S$ two-pion exchange component, which is partially absorbed 
in the $\Lambda N$ and $\Lambda NN$ LO contact LECs. Short-range 
$K$ and $K^\ast$ meson exchanges induce a mild $\Lambda N$ tensor 
force~\cite{gsd71,mgdd85}, the weakness of which is confirmed in shell-model 
studies of observed $p$-shell $\Lambda$ hypernuclear spectra~\cite{mil12}. 
Such momentum-dependent interaction terms which appear at subleading order 
in $\nopieft$ power counting, need to be introduced systematically in future 
applications to $p$-shell hypernuclei. 

Apart from the two-body contact terms that are specified here by $NN$ and 
$\Lambda N$ spin-singlet and triplet scattering lengths, amounting to four 
low-energy constants (LECs), the theory uses additionally four three-body 
LECs: a pure $NNN$ LEC fitted to $B(^3$H) and three $\Lambda NN$ LECs 
associated with the three possible $s$-wave $\Lambda NN$ systems, 
of which only \lamb{3}{H}($I$=0,\,$J^P$=${\frac{1}{2}}^+$) is bound. 
Therefore, on top of fitting its binding energy, the binding energies 
of \lamb{4}{H}$_{\rm g.s.}$($I$=${\frac{1}{2}}$,\,$J^P$=$0^+$) and of   
\lamb{4}{H}$_{\rm exc}$($I$=$\frac{1}{2}$,\,$J^P$=$1^+$) are also fitted. 
The fitted LECs are used then, for a sequence of $\lambda$ cutoff values, 
to evaluate the binding energies of $^4$He and \lamb{5}{He}. Remarkably, 
$B(^4$He) is reproduced well in the renormalization scale invariance limit 
$\lambda\to\infty$.

\begin{figure}[htb] 
\begin{center} 
\includegraphics[width=0.48\textwidth]{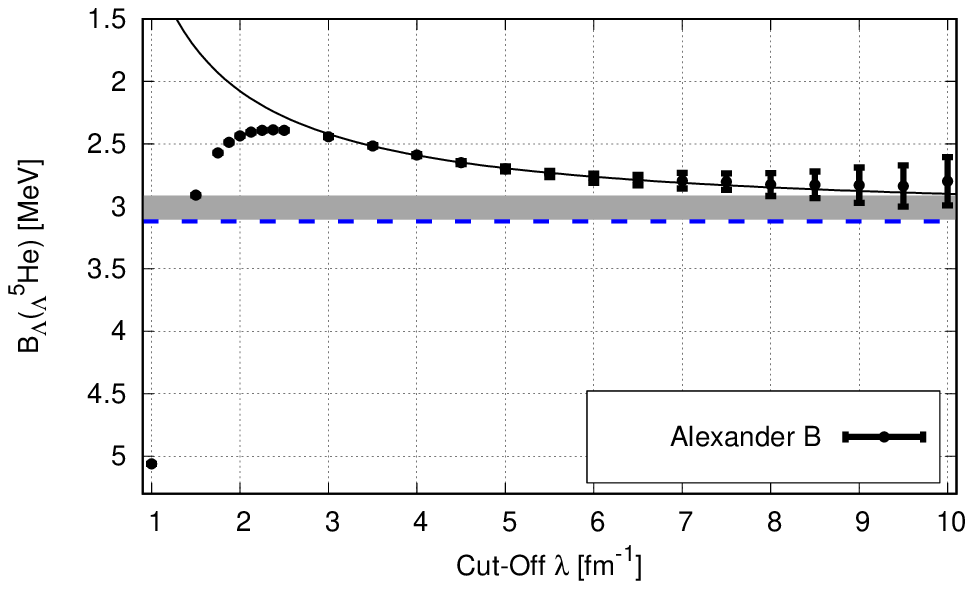}
\includegraphics[width=0.48\textwidth]{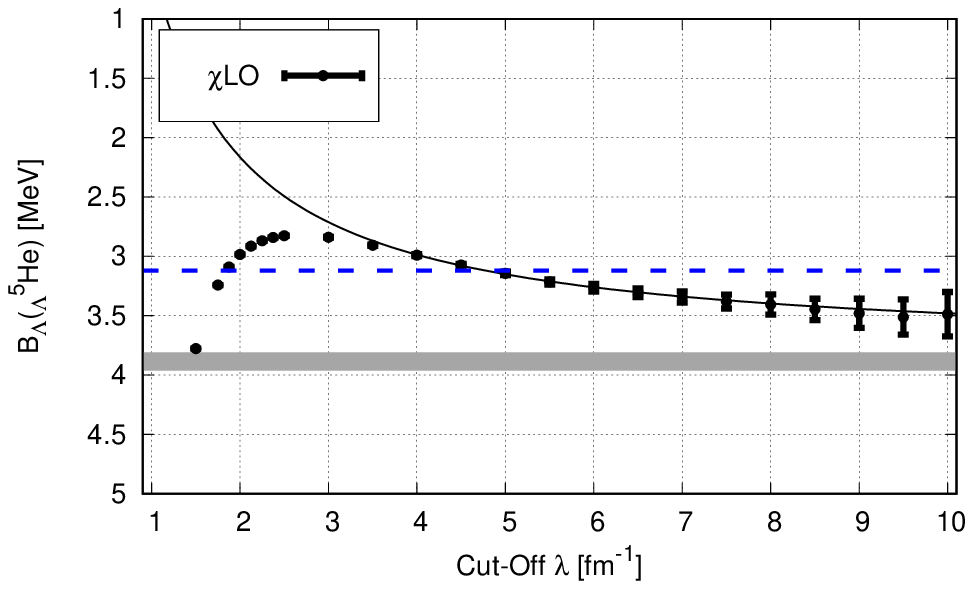}
\caption{$B_{\Lambda}$(\lamb{5}{He}) as a function of $\lambda$ in 
\nopieft~calculations with $\Lambda N$ input from $\Lambda p$ scattering 
experiments~\cite{Alex68} (left panel) and from a LO $\chi$EFT 
model~\cite{phm06} (right panel). Solid lines mark a fit $a+b/\lambda$ 
for $\lambda\geq 4$~fm$^{-1}$. Horizontal bands mark $\lambda\to\infty$ 
extrapolation uncertainties. Dashed horizontal lines mark the value 
$B_{\Lambda}^{\rm exp}(_{\Lambda}^5$He)=3.12$\pm$0.02~MeV. Figure adapted 
from Ref.~\cite{cbg18}.} 
\label{fig:cbg} 
\end{center} 
\end{figure} 

The \nopieft~approach was applied in SVM few-body calculations of $s$-shell 
hypernuclei, using several models of the $\Lambda N$ scattering lengths. 
The resulting $\Lambda$ separation energy values $B_{\Lambda}(_\Lambda^5$He) 
are shown in Fig.~\ref{fig:cbg} for two such models as a function of 
the cutoff $\lambda$. Common to all $\Lambda N$ models, the calculated 
$B_{\Lambda}(_\Lambda^5$He) values switch from about 2--3~MeV overbinding 
at $\lambda$=1~fm$^{-1}$ to less than 1~MeV underbinding between 
$\lambda$=2 and 3~fm$^{-1}$, and smoothly varying beyond, approaching 
a finite (renormalization scale invariance) limit at $\lambda\to\infty$. 
A reasonable choice of {\it finite} cutoff values in the present case is 
between $\lambda$$\approx$1.5~fm$^{-1}$, which marks the \nopieft~breakup 
scale of 2$m_{\pi}$, and 4~fm$^{-1}$, beginning at which the detailed dynamics 
of vector-meson exchanges may require attention. We note that for $\lambda
\gtrsim 1.5$~fm$^{-1}$ all of the three $\Lambda NN$ state components are 
repulsive, as required to avoid Thomas collapse. Recent LO $\chi$EFT 
calculations~\cite{wr16} using induced $YNN$ repulsive contributions 
suggest that the $s$-shell overbinding problem extends to the $p$ shell. 
Interestingly, shell-model studies~\cite{mil12} reproduce satisfactorily 
$p$-shell ground-state $B_{\Lambda}$ values, essentially by using 
$B_{\Lambda}^{\rm exp}(_{\Lambda}^5$He) for input, except for the relatively 
large difference of about 1.8~MeV between $B_{\Lambda}({_{\Lambda}^9}$Li) 
and $B_{\Lambda}({_{\Lambda}^9}$Be). In fact, it was noted long ago that 
strongly repulsive $\Lambda NN$ terms could settle it~\cite{gal67}. It would 
be interesting to test the $\Lambda NN$ interaction terms derived here in 
future shell-model studies and, perhaps, also in NS matter calculations such 
as by Lonardoni {\it et al.}~\cite{lpg14} that are geared to resolve the 
`hyperon puzzle'~\cite{llgp15}.

\section{Onset of binding in $\Lambda\Lambda$ hypernuclei}
\label{sec:LL} 

The second topic reviewed here is the onset of binding $\Lambda\Lambda$ 
hypernuclei~\cite{csbgm19}, using a methodology similar to that used for the 
previous topic. Reliable data on $\Lambda\Lambda$ hypernuclei are scarce: the 
Nagara event~\cite{nagara01,ahn13} is perhaps the {\it only} $\Lambda\Lambda$ 
hypernucleus determined unambiguously, identified as \lamblamb{6}{He}, with 
two more $\Lambda\Lambda$ hypernuclei, \lamlam{10}{Be} and \lamlam{13}{B}, 
that are also generally accepted~\cite{gm11}. 
The \lamblamb{6}{He} datum $\Delta B_{\Lambda\Lambda}$(\lamblamb{6}{He})=$
B_{\Lambda\Lambda}$(\lamblamb{6}{He})$-$2$B_{\Lambda}$(\lamb{5}{He})=0.67$
\pm$0.17~MeV~\cite{ahn13} serves as a constraint, assuming also that the 
low-energy $\Lambda\Lambda$ interaction is weaker than the $\Lambda N$ 
interaction. The onset of $\Lambda\Lambda$ hypernuclear binding is then found 
at the isodoublet \lamblamb{5}{H}--\lamblamb{5}{He}, with a \lamblamb{4}{H} 
bound state not definitively excluded. 

\begin{figure}[htb] 
\begin{center} 
\includegraphics[width=0.48\textwidth,height=6cm]{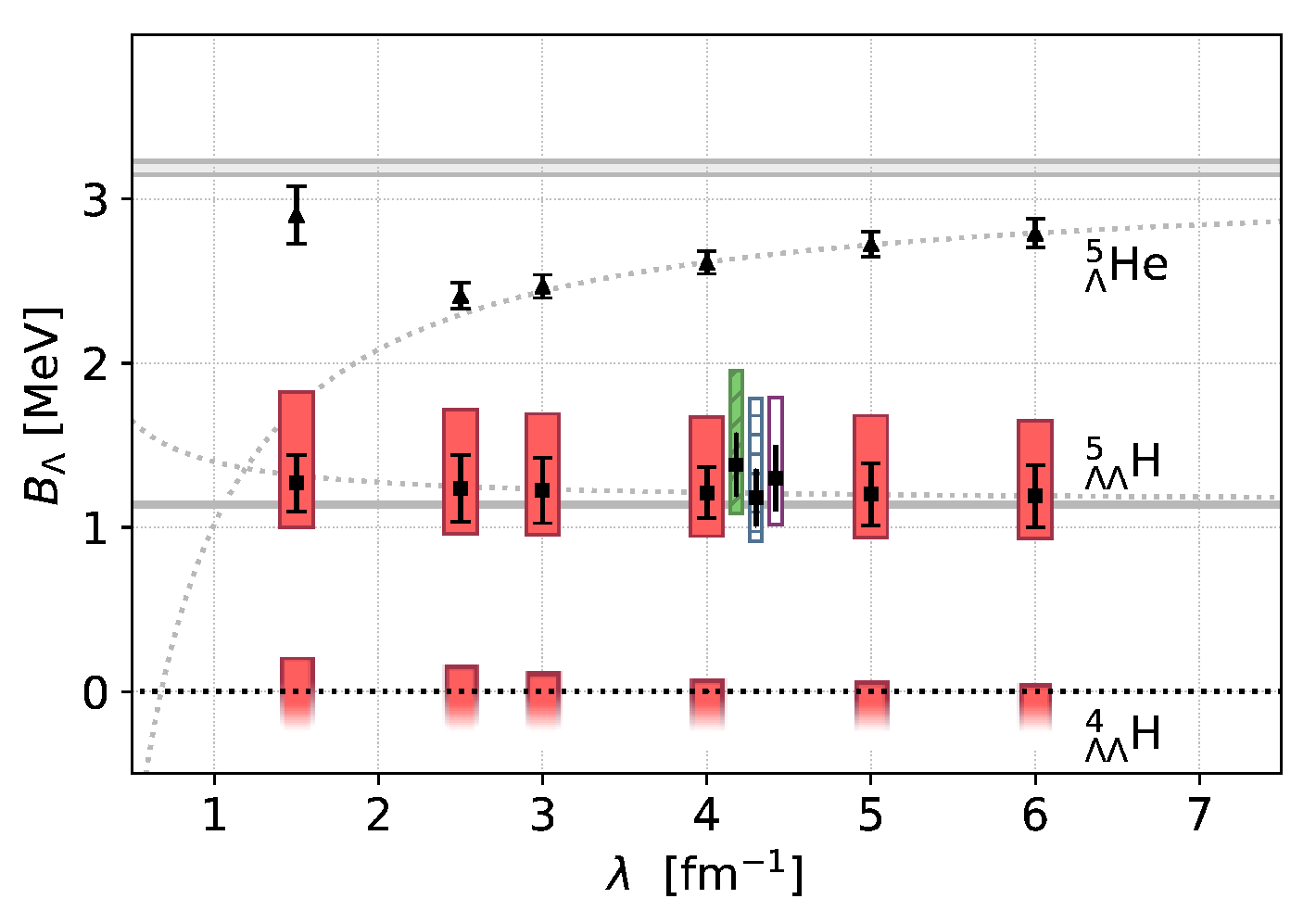}
\includegraphics[width=0.48\textwidth]{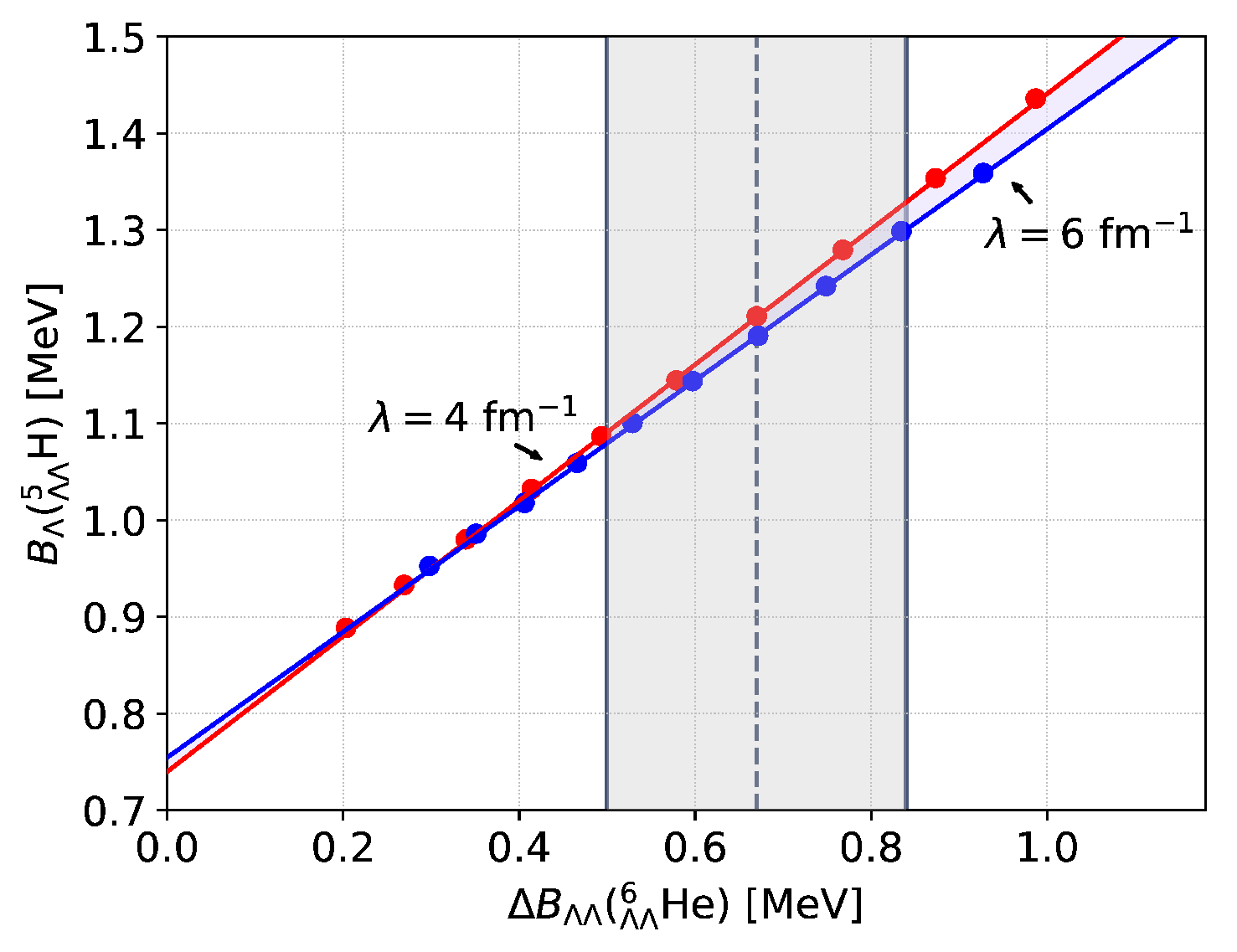} 
\caption{Left: $B_{\Lambda}$(\lamb{5}{He},~\lamblamb{4}{H},~\lamblamb{5}{H}) 
from \nopieft~calculations~\cite{cbg18,csbgm19}. Error bars (in black) reflect 
given uncertainties in the $^3_\Lambda$H, $^4_\Lambda$H, $^4_\Lambda$H$^\ast$ 
and $^{~~6}_{\Lambda\Lambda}$He binding-energy input data, and rectangles 
(in red) arise from varying $a_{\Lambda\Lambda}$ between $-$0.5 to $-$1.9~fm. 
Dotted lines show extrapolations to $\lambda\to\infty$ limits marked by gray 
horizontal bands. 
Right: Tjon lines relating calculated $B_{\Lambda}$(\lamblamb{5}{H}) values to 
$\Delta B_{\Lambda\Lambda}$(\lamblamb{6}{He}) values beyond its experimental 
datum marked by vertical straight lines.} 
\label{fig:LL5H} 
\end{center} 
\end{figure} 

$\Lambda$ separation energy values $B_\Lambda(^{~~5}_{\Lambda\Lambda}$H) from 
\nopieft~calculations~\cite{csbgm19} are shown in Fig.~\ref{fig:LL5H}. Several 
representative values of the $\Lambda\Lambda$ scattering length were used, 
spanning a broad range of values suggested by analyses of $\Lambda\Lambda$ 
correlations observed recently in relativistic heavy-ion collisions and 
by analyzing the KEK-PS E522~\cite{Yoon07} invariant mass spectrum in the 
reaction $^{12}$C($K^-,K^+)\Lambda\Lambda X$ near the $\Lambda\Lambda$ 
threshold; see Ref.~\cite{csbgm19} for detailed references. Here the choice of 
$a_{\Lambda\Lambda}$ determines the one $\Lambda\Lambda$ LEC required at LO, 
while the $\Lambda\Lambda N$ LEC was fitted to the $\Delta B_{\Lambda\Lambda}
(^{~~6}_{\Lambda\Lambda}$He)=0.67$\pm$0.17~MeV datum~\cite{ahn13}. Most 
calculations were made using the Alexander[B] $\Lambda N$ model with 
scattering lengths $a_{s,t}$=$-1.8,-1.6$~fm~\cite{Alex68}, but for cutoff 
$\lambda$=4~fm$^{-1}$ three other $\Lambda N$ interaction models from 
Ref.~\cite{cbg18} were also used, demonstrating that the $\Lambda N$ model 
dependence is rather weak when it comes to double-$\Lambda$ hypernuclei, 
provided $B_\Lambda$ values of single-$\Lambda$ hypernuclei for $A<5$ are 
fitted to generate the necessary $\Lambda NN$ LECs. Calculated values of 
$B_\Lambda(^5_\Lambda$He) are also shown in the figure as a check. One 
observes that $^{~~5}_{\Lambda\Lambda}$H comes out particle stable over 
a broad range of cutoff values used in the calculations. This is not the 
case for $^{~~4}_{\Lambda\Lambda}$H which comes out unbound with respect 
to $^3_\Lambda$H for most of the permissible parameter space. Finally, 
`Tjon line' correlations~\cite{Tjon75} found between $B_\Lambda(^{~~5}_{
\Lambda\Lambda}$H) and $B_\Lambda(^{~~6}_{\Lambda\Lambda}$He), when $\Delta 
B_{\Lambda\Lambda}(^{~~6}_{\Lambda\Lambda}$He) is varied within and also 
outside of its reported error-bar values, are demonstrated in the right panel 
of Fig.~\ref{fig:LL5H}. Such correlations were noted already in old Faddeev 
calculations, e.g. Ref.~\cite{fg02}. 

\begin{table}[htb] 
\caption{$\Lambda$ separation energies $B_\Lambda(^{~~A}_{\Lambda\Lambda}$Z) 
for $A$=3--6, calculated using $a_{\Lambda\Lambda}$=$-0.8$~fm, cutoff 
$\lambda$=4~fm$^{-1}$ and the Alexander[B] $\Lambda N$ interaction
model~\cite{Alex68}, see text.} 
\vspace{5pt} 
{\renewcommand{\arraystretch}{1.1}} 
\begin{center} 
\begin{tabular}{lccccc} 
\br 
Constraint (MeV) & $^{~~3}_{\Lambda\Lambda}$n & $^{~~4}_{\Lambda\Lambda}$n & 
$^{~~4}_{\Lambda\Lambda}$H & $^{~~5}_{\Lambda\Lambda}$H & $^{~~6}_{\Lambda
\Lambda}$He \\ 
\mr 
$\Delta B_{\Lambda\Lambda}(^{~~6}_{\Lambda\Lambda}$He)=$\underline{0.67}$ 
& -- & -- & -- & 1.21 & 3.28 \\ 
$B_{\Lambda}(^{~~4}_{\Lambda\Lambda}$H)=$\underline{0.05}$ & -- & -- & 0.05 
& 2.28 & 4.76 \\ 
$B(^{~~4}_{\Lambda\Lambda}$n)=$\underline{0.10}$ & -- & 0.10 & 0.86 & 4.89 
& 7.89 \\ 
$B(^{~~3}_{\Lambda\Lambda}$n)=$\underline{0.10}$ & 0.10 & 15.15 & 18.40 
& 22.13 & 25.66 \\ 
\br 
\end{tabular} 
\end{center} 
\label{tab:LL} 
\end{table} 

To make $^{~~4}_{\Lambda\Lambda}$H particle stable one may reduce the 
repulsive $\Lambda\Lambda N$ LEC from its value constrained by the $^{~~6}_{
\Lambda\Lambda}$He datum, using representative values for $a_{\Lambda\Lambda}$ 
and the cutoff $\lambda$ for which $^{~~4}_{\Lambda\Lambda}$H was found 
particle unstable. According to the first two rows in Table~\ref{tab:LL}, 
this will overbind $^{~~6}_{\Lambda\Lambda}$He by $\approx$1.5~MeV. Reducing 
further the $\Lambda\Lambda N$ LEC one binds the neutral systems, first 
$^{~~4}_{\Lambda\Lambda}$n (third row) and then $^{~~3}_{\Lambda\Lambda}$n 
(fourth row), at a price of overbinding further $^{~~6}_{\Lambda\Lambda}$He. 
These results strongly suggest that the $A=3,4$ neutral $\Lambda\Lambda$ 
hypernuclei are unbound within a large margin.

\section{\lamb{3}{H} and \lamb{3}{n} lifetime puzzles} 
\label{sec:L3H} 

The third topic discussed here is the hypertriton lifetime puzzle: why is the 
lifetime of the loosely bound \lamb{3}{H} shorter than the free-$\Lambda$ 
lifetime $\tau_{\Lambda}$ by as much as $\approx (30\pm 8)\%$, as suggested 
in recent measurements using relativistic heavy ion collisions to produce 
light nuclei, anti-nuclei and hyperfragments~\cite{BMD18} such as \lamb{3}{H}? 
Also discussed briefly is the lifetime expected for the questionable 
\lamb{3}{n}~\cite{Rap13,Saito16}. This 3-body hypernucleus has been found 
unanimously particle unstable in several few-body calculations cited below, 
including Ref.~\cite{cbg18} discussed in this report.  

Measurements of the \lamb{3}{H} lifetime in emulsion or bubble-chamber 
experiments during the 1960s and early 1970s gave conflicting and puzzling 
results. Particularly troubling appeared a conference report by Block 
{\it et al.} claiming a lifetime of $\tau$(\lamb{3}{H})=(95$^{+19}_{-15}
$)~ps~\cite{Block62}, to be compared with a free $\Lambda$ lifetime 
$\tau_\Lambda$=(236$\pm$6)~ps measured in the same He chamber~\cite{Block63}. 
Given the loose $\Lambda$ binding, $B_\Lambda$(\lamb{3}{H})=0.13$\pm$0.05~MeV, 
it was anticipated that $\tau$(\lamb{3}{H})$\approx$$\tau_\Lambda$, as argued 
by Rayet and Dalitz~(RD)~\cite{RD66} using a closure-approximation approach 
for \lamb{3}{H}$_{\rm g.s.}({\frac{1}{2}}^+)$ decay. 
 
\begin{table}[htb] 
\caption{\lamb{3}{H}$_{\rm g.s.}({\frac{1}{2}}^+)$ decay rate calculated in 
units of the free $\Lambda$ decay rate $\Gamma_\Lambda$ and listed in a year 
of publication order. The first row lists results for plane-wave pions, 
disregarding pion final state interaction (FSI) contributions which are listed 
in the second row. A calculated nonmesonic decay rate contribution of 0.017 
from Ref.~\cite{Golak97} was added uniformly in obtaining the total decay 
rates listed in the third, last row.} 
\vspace{5pt}
{\renewcommand{\arraystretch}{1.1}}
\begin{center} 
\begin{tabular}{ccccc}
\br 
$\Gamma$(\lamb{3}{H}) model & 1966~\cite{RD66} & 1992~\cite{Con92} & 
1998~\cite{Fad98} & 2019~\cite{GG18} \\  
\mr  
Without pion FSI & 1.05 & 1.12 & 1.01 & 1.11 \\ 
Pion FSI contribution & $-$0.013 & -- & -- & 0.11 \\  
Total & 1.05 & 1.14 & 1.03 & 1.23 \\  
\br  
\end{tabular}
\end{center} 
\label{tab:L3H} 
\end{table} 

Table~\ref{tab:L3H} lists \lamb{3}{H}$_{\rm g.s.}({\frac{1}{2}}^+)$ decay 
rate values calculated by RD and in several subsequent solid calculations, all 
reaching similar results. Claims for large departures from the free $\Lambda$ 
value are, as a rule, incorrect or irreproducible. The RD methodology was also 
used by Congleton~\cite{Con92} and by Gal and Garcilazo~(GG)~\cite{GG18}, with 
the latter one solving appropriate three-body Faddeev equations to produce 
a \lamb{3}{H} wavefunction. The Kamada {\it et al.} calculation~\cite{Fad98}, 
while also solving Faddeev equations for the \lamb{3}{H} wavefunction, 
accounted microscopically for the outgoing 3$N$ phase space and FSI, thereby 
doing without a closure approximation. Pion FSI was considered only in two of 
these works, with differing results: (i) repulsion, weakly reducing $\Gamma
$(\lamb{3}{H}) in RD; and (ii) attraction, moderately enhancing it in GG. 
The latter result is supported by the $\pi^-$-atom $1s$ level {\it attractive} 
shift observed in $^3$He~\cite{Schwanner84}. It is remarkable that \lamb{3}{H} 
decay is the only light hypernucleus decay where the low-energy pion $s$-wave 
FSI is expected to be attractive. The decays of \lamb{4}{H}, \lamb{4}{He} and 
\lamb{5}{He} involve pion-$^4$He FSI which is known from the $\pi^-$ atomic 
$^4$He $1s$ level shift to be repulsive~\cite{Backenstoss74}. 

\begin{figure}[!t]  
\begin{center} 
\includegraphics[width=\textwidth]{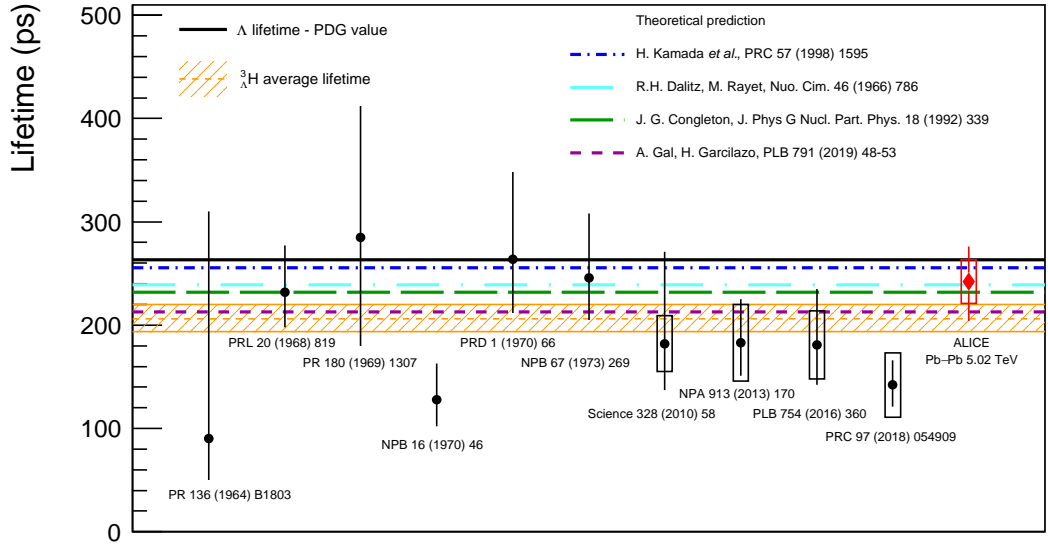} 
\caption{Compilation of measured \lamb{3}{H} lifetime values, plotted in 
chronological order. The five most recent ones are from relativistic heavy ion 
experiments. Shown by a solid (black) horizontal line is the free-$\Lambda$ 
lifetime, with the world average of measured \lamb{3}{H} lifetimes marked by 
an orange band, and calculated lifetimes listed in Table~\ref{tab:L3H} marked 
by dashed horizontal lines. Figure courtesy of Benjamin 
D\"{o}nigus~\cite{BMD18}.} 
\label{fig:L3H} 
\end{center} 
\end{figure} 

Renewed interest in the \lamb{3}{H} lifetime problem arose by recent 
measurements of $\tau$(\lamb{3}{H}) in relativistic heavy ion experiments 
marked in Fig.~\ref{fig:L3H} (STAR~\cite{STAR10}, HypHI~\cite{HypHI13}, 
ALICE~\cite{ALICE16}, STAR~\cite{STAR18} and ALICE~\cite{ALICE19}) reporting 
values shorter by (28$\pm$8)\% than $\tau_\Lambda$=(263$\pm$2)~ps~\cite{BMD18}. 
While enhancement of the free $\Lambda$ decay rate by up to $\approx$20\% 
is theoretically conceivable relying on the new GG calculation~\cite{GG18} as 
recorded in the last column of Table~\ref{tab:L3H}, it appears inconceivable 
at present to reproduce a 30\% or even larger enhancement suggested by some 
of the recent heavy-ion experiments. Note however that the most recent ALICE 
lifetime result~\cite{ALICE19} is compatible within errors with the listed 
calculated values and also with $\tau_\Lambda$. 

\newpage 

Before closing this section I wish to make a few remarks on \lamb{3}{n}, 
conjectured by the HypHI GSI Collaboration~\cite{Rap13} to be bound, while 
unbound in recent theoretical calculations~\cite{GV14,Hiyama14,GG14}. 
In \lamb{3}{n} decays induced by $\Lambda\to p+\pi^-$, where the \lamb{3}{n} 
neutrons are spectators, the \lamb{3}{n}~$\to (pnn) + \pi^-$ weak decay rate 
is given in the closure approximation essentially by the $\Lambda\to p+\pi^-$ 
free-space weak-decay rate, whereas in $\Lambda\to n+\pi^0$ induced decays 
the production of a third low-momentum neutron is suppressed by the Pauli 
principle, and this \lamb{3}{n} weak decay branch may be disregarded up to 
perhaps a few percents. Hence $\Gamma({_\Lambda^3}{\rm n})/\Gamma_\Lambda
\approx 1.114\times 0.641=0.714$~\cite{GG18}, where the factor 1.114 follows 
from the difference between the recoil energies in the corresponding phase 
space factors, and the factor 0.641 is the free-space $\Lambda\to p+\pi^-$ 
fraction of the total $\Lambda\to N+\pi$ weak decay rate, giving rise to an 
estimated \lamb{3}{n} lifetime of $\tau$(\lamb{3}{n})$\approx 368$~ps, which 
should hold up to a few percent contribution from the $\pi^0$ decay branch. 
This lifetime is considerably longer than 181${^{+30}_{-24}}\pm$25~ps or 
190${^{+47}_{-35}}\pm$36~ps deduced from the $nd\pi^-$ and $t\pi^-$ alleged 
decay modes of \lamb{3}{n}~\cite{Rap13,Saito16}, providing a strong argument 
against the conjectured stability of \lamb{3}{n}.

\ack 
I thank the Organizers of the INPC-2019 for their kind hospitality. I am also 
indebted to my good colleagues Nir Barnea, Lorenzo Contessi, Eli Friedman, 
Humberto Garcilazo, Ji\v{r}\'{i} Mare\v{s} and Martin Sch\"{a}fer for insights 
gained through our long-standing collaborations.

\end{document}